\documentclass[aps,prb,12pt,amsmath,amssymb,notitlepage]{revtex4-1}

\usepackage{natbib}
\usepackage{graphicx}
\usepackage{amsmath}
\usepackage{amssymb}
\usepackage[usenames]{color}

\newcounter{lastnote}

\begin{document} 

\title{Multi-terminal Josephson junctions as topological materials}

\author{Roman-Pascal~Riwar$^*$, Manuel~Houzet, Julia~S.~Meyer}
\affiliation{Univ. Grenoble Alpes, INAC-SPSMS, F-38000 Grenoble, France,\\
CEA, INAC-SPSMS, F-38000 Grenoble, France
}
\author{Yuli~V.~Nazarov}
\affiliation{Kavli Institute of NanoScience, Delft University of Technology,\\ Lorentzweg 1, NL-2628 CJ, Delft, The Netherlands.}

\begin{abstract}
Topological materials and their unusual transport properties are now at the focus of modern experimental and theoretical research. Their topological properties arise from the bandstructure determined by the atomic composition of a material and as such are difficult to tune and naturally restricted to $\leq3$ dimensions. Here we demonstrate that $n$-terminal Josephson junctions with conventional superconductors may provide a straightforward realization of tunable topological materials in $n-1$ dimensions. For $n\geq 4$, the Andreev subgap spectrum of the junction can accommodate Weyl singularities in the space of the $n-1$ independent superconducting phases, which play the role of bandstructure quasimomenta. 
The presence of these Weyl singularities enables topological transitions that are manifested experimentally as changes of the quantized transconductance between two voltage-biased leads, the quantization unit being $2e^2/(\pi\hbar)$. \\ \\
$^*$To whom correspondence should be addressed; E-mail: roman-pascal.riwar@cea.fr
\end{abstract}

\date{\today}

\maketitle 

Josephson junctions created by coupling \textit{two} superconductors through a weak link have been studied extensively for many years~\cite{Josephson1962,Golubov2004,Devoret2004}. The current across a Josephson junction yields information about the Andreev bound states (ABS) forming at the junction~\cite{Beenakker1991,DellaRocca2007,Pillet2010,Bretheau2013b}. In turn, the ABS spectrum is determined by the properties of the junction and the superconducting leads.
For instance, if the leads are topologically non-trivial, the Josephson effect may be used to probe these topological properties.
In particular, the $4\pi$-periodicity of the supercurrent indicates the presence of topologically protected zero-energy Majorana states~\cite{Kitaev2001,Fu2009,Alicea2012,Badiane2013}, which may arise in  1D spinless p-wave superconductors, semiconductor nanowires with proximity-induced superconductivity, or at the surface of bulk materials.

In this paper, we show that \textit{multi}-terminal Josephson junctions may be topologically nontrivial
even if the superconducting leads are topologically trivial and no exotic materials are used to make the junction. 
Thus, the junction itself may be regarded as an artificial topological material, which displays Weyl singularities.

The rapidly growing field of 3D Weyl semimetals deals with a bandstructure that exhibits conical energy gap closings: Weyl points~\cite{Herring1937,Murakami2007,Wan2011,Burkov2011,Hosur2013}. Unlike the Dirac point in graphene~\cite{CastroNeto2009} that may
be gapped out through an appropriate coupling, isolated Weyl points are topologically protected. They
can be regarded as monopoles with $\pm$ charge.
A topological invariant -- the Chern number -- defined on a surface in momentum space characterizes the total charge of the monopoles it encloses.
These monopoles give rise to many unusual features, such as chiral edge states and associated surface Fermi arcs~\cite{Wan2011}. Topological protection guarantees that the only way to induce a gap is either to annihilate two Weyl points of opposite charge by bringing them together, or to couple two cones at a finite distance in momentum space through breaking of momentum conservation~\cite{Hosur2013}.

Here, we demonstrate Weyl singularities in multi-terminal junctions with conventional s-wave superconducting leads, namely when the energy of the lowest ABS goes to zero at certain values of the superconducting phases such that the gap in the spectrum closes. Below, we also show that their topological property can be easily probed by the transconductance between two voltage-biased leads, which is proportional to the Chern number.

\begin{figure}
\begin{center}
\includegraphics[scale=0.8]{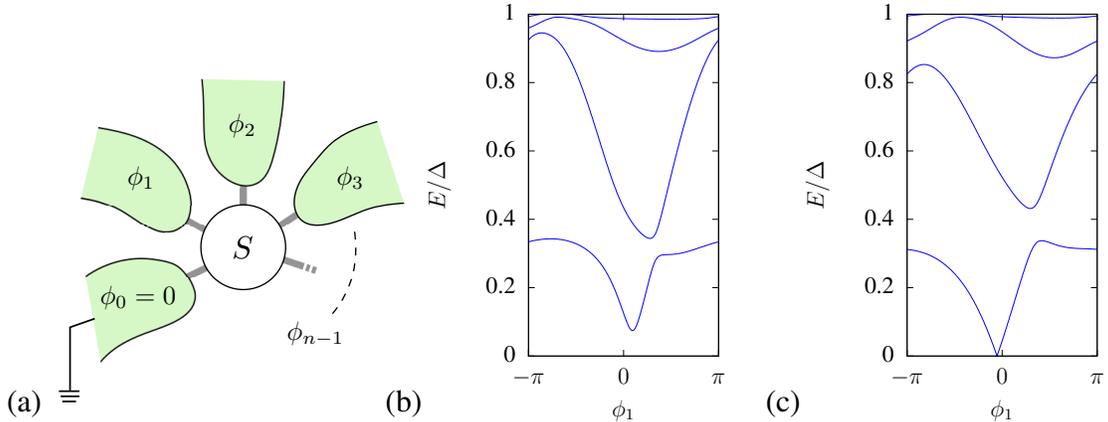}
\end{center}
\caption{\textbf{General setup of the multi-terminal junction, and examples of typical ABS spectra.} a) The superconducting leads with phases $\phi_\alpha$, $\alpha={0,\ldots,n-1}$, are connected through a scattering region described by the scattering matrix $S$. b) Generic ABS energy spectrum versus $\phi_1$, away from a Weyl singularity. c) ABS energy spectrum versus $\phi_1$, where the other phases are tuned to a Weyl singularity. Note the gap closing. }\label{fig_sketch}
\end{figure}

We consider a junction with $n$ superconducting leads connected through a scattering region (see Fig.~\ref{fig_sketch}a). The leads $\alpha=\{0,1,\ldots,n-1\}$ have the same gap $\Delta$, though they may differ in the phase of the superconducting order parameter, $\phi_\alpha$.
Due to gauge invariance, only $n-1$ phases are independent, hence we may set $\phi_0=0$. Likewise we choose to focus on a short scattering region, characterized by an energy- (and spin-) independent scattering matrix $S$ in the basis of $N=\sum_\alpha N_\alpha$ transport channels, $N_\alpha$ being the number of channels in contact $\alpha$. We also assume time-reversal symmetry, such that $S^{T}=S$. 

The junction hosts a set of spin-degenerate Andreev bound states, indexed by $k$, whose energies $E_k\geq0$ are determined from the equation~\cite{Beenakker1991}
\begin{equation}\label{eq_Beenakker}
{\rm det} \left[ 1 - e^{-2i\chi} S e^{i \hat{\phi}} S^* e^{-i \hat{\phi}}\right]=0\ ,
\end{equation}
where $\chi = \arccos(E/\Delta)$, and $e^{i \hat{\phi}}$ is a diagonal matrix that assigns to each channel the phase factor of the corresponding terminal. 
The spin-degenerate ABS energies $E_k(\hat{\phi})$ (see Fig.~\ref{fig_sketch}b) are periodic in all phases with a period $2\pi$.
The total phase-dependent energy of the junction reads $E=\sum_{k\sigma} (n_{k\sigma} - 1/2) E_k$, $n_{k\sigma} = 0,1$ being the occupation of the state $k$ with spin $\sigma$. 

If the gap closes at certain values of the phases, $\hat{\phi}^{(0)}$, the spectrum is doubly degenerate at $E=0$. We can then describe the lowest energy band in the vicinity of the gap closing by the two-by-two Weyl Hamiltonian (Supplementary Discussion),
\begin{equation}\label{eq_HWeyl}
H_\text{W}=\sum_{i=x,y,z}h_i\tau_i\ ,\quad h_i=\sum_\alpha \delta\phi_\alpha M_{\alpha i}\ ,
\end{equation}
with the Pauli matrices $\tau_{x,y,z}$ in the basis of the two degenerate states corresponding to eigenvalue $E=0$.
The fields $h_i$ depend linearly on $\delta\hat{\phi}=\hat{\phi}-\hat{\phi}^{(0)}$ through the real matrix $M$.

The form of the Weyl Hamiltonian in Eq.~\eqref{eq_HWeyl} indicates that we need at least three parameters to tune the system to the degeneracy point, $h_i=0$.
Thus, for 4 terminals with 3 independent phases, the Weyl singularities appear as points in the 3D phase space. For 5 terminals, the Weyl singularities occur in general as 1D curves in the 4-dimensional space of phases. This opens up the possibility to realize a topological material in arbitrary dimensions.
Note that such multi-terminal Josephson junctions cannot be characterized by means of the standard ``periodic table'' of topological semimetals~\cite{Matsuura2013,Zhao2013}, due to the distinct behaviour of the quasimomenta $\hat{\phi}$ under particle-hole symmetry. A proper classification could be envisioned along the lines of Ref.~\cite{Zhang2014}.

The topology of the junction can be characterized by a set of Chern numbers. A Chern number may be defined in the 2D subspace of two phases $\phi_\alpha$ and $\phi_\beta$, through the local Berry curvature of the ABS. The Berry curvature for the bound state $k$ with spin $\sigma$ is related to its wave function $|\varphi_{k\sigma}\rangle$ through
\begin{equation}
B_k^{\alpha\beta} \equiv -2 \text{Im} \left\langle\frac{\partial\varphi_{k\sigma}}{\partial \phi_\alpha}\left|\frac{\partial\varphi_{k\sigma}}{\partial \phi_\beta}\right\rangle\right..
\end{equation}
Note that $B_k^{\alpha\beta}$ does not depend on spin (Supplementary Discussion). The total Berry curvature of the many-body superconducting state then reads $B^{\alpha\beta} = \sum_{k\sigma} (n_{k\sigma} - 1/2) B^{\alpha\beta}_k$, in analogy to the expression for the energy. For fixed occupations $n_{k\sigma}=0,1$, the integral of the Berry curvature over the elementary cell yields an integer, the Chern number $C^{\alpha\beta}=\int_{-\pi}^\pi\int_{-\pi}^\pi d\phi_\alpha d\phi_\beta\,B^{\alpha\beta}/(2\pi)$.

Since the Weyl singularities appear as points in the 3D phase space, a third phase $\phi_\gamma$ may be used to tune the system through Weyl points, thus changing the Chern number. We see that a given state $k$ only contributes to $B^{\alpha\beta}$ when it has an even occupation number, $n_{k\uparrow}=n_{k\downarrow}$. The topological signature is therefore robust provided the fermion parity is preserved, a fact which has also been pointed out  for other topological systems realized in superconductors~\cite{Fu2009}. Quasiparticle poisoning, by which a single quasiparticle enters a bound state, enables transitions between even and odd ABS occupation~\cite{Zgirski2011}, and can therefore affect the signal. In the following, we focus on the ground state, $n_{k\sigma}=0$ for all $k$ and $\sigma$, where the Chern number may be nontrivial.
 
Importantly, the current response of the junction with slowly varying phases reveals the Chern number. Biasing lead $\beta$ with a voltage $eV_{\beta}\ll \Delta$ gives rise to the instant current to contact $\alpha$ (Supplementary Discussion)
\begin{equation}
I_\alpha(t)=\frac{2e}{\hbar}\frac{\partial E}{\partial\phi_\alpha}-2e\dot{\phi}_\beta B^{\alpha\beta},
\end{equation}
where $\dot{\phi}_{\beta} = 2e V_{\beta}/\hbar$. The first term corresponds to the adiabatic current and the second term is the first order correction in the phase velocity. Let us now apply constant voltages to two leads. For incommensurate voltages, the two phases uniformly sweep the elementary cell. In the dc limit, the adiabatic current contribution then averages out, and the Berry curvature is replaced by its average value.
Thus, we find that the dc current is linear in the voltages, and the transconductance is defined by the Chern number
\begin{equation}\label{eq_transcond}
\overline{I}_\alpha=G^{\alpha\beta} V_\beta \quad {\rm with} \quad G^{\alpha\beta} = -\frac{2e^2}{\pi\hbar} C^{\alpha\beta}.
\end{equation} 
Equation~\eqref{eq_transcond} shows that multi-terminal junctions exhibit a dc current response typical for the quantum Hall effect, 
although based on different physics. The transconductance quantum is four times bigger than in the quantum Hall effect, which can be traced to the $2e$ charge of the superconducting Cooper pairs. To extract the small dc signal, the averaging time needs to be sufficiently long. The relevant time scale is determined by the low-frequency current noise (Supplementary Discussion). Note that quantum Hall-like conductance quantization has also been proposed in superconducting devices with finite charging energy and hosting quantum phase slips~\cite{Hriscu2013}.
Furthermore, superconducting junctions with a gate-tunable charging energy may realize topologically-protected discrete charge pumping~\cite{Leone2008}.

We now focus on a 4-terminal junction and investigate the energy spectrum as a function of the three independent phases $\phi_{1,2,3}$. As mentioned above, such a 3D bandstructure may host Weyl points with positive or negative topological charge. The Nielsen-Ninomiya theorem~\cite{Nielsen1981} implies that the total topological charge of the system is zero, such that the number of Weyl points is always even. Furthermore, time-reversal invariance corresponds to a mapping from $\hat{\phi}$ to $-\hat{\phi}$, hence a Weyl point at $\hat{\phi}^{(0)}$ has a counterpart at $-\hat{\phi}^{(0)}$ with the same topological charge. Thus, Weyl points come in groups of 4.

In the simplest case, where each contact contains only one channel, the system may realize 0 or 4 Weyl points, corresponding to a topologically trivial or nontrivial 3D material, respectively. If a scattering matrix yielding Weyl points is found, small changes in the scattering matrix only modify their position, but cannot gap them. Namely, as the Weyl points carry a topological charge, individual Weyl points are stable and annihilation is possible only when two Weyl points with opposite charges coincide.

\begin{figure}
\centering
\includegraphics[scale=0.8]{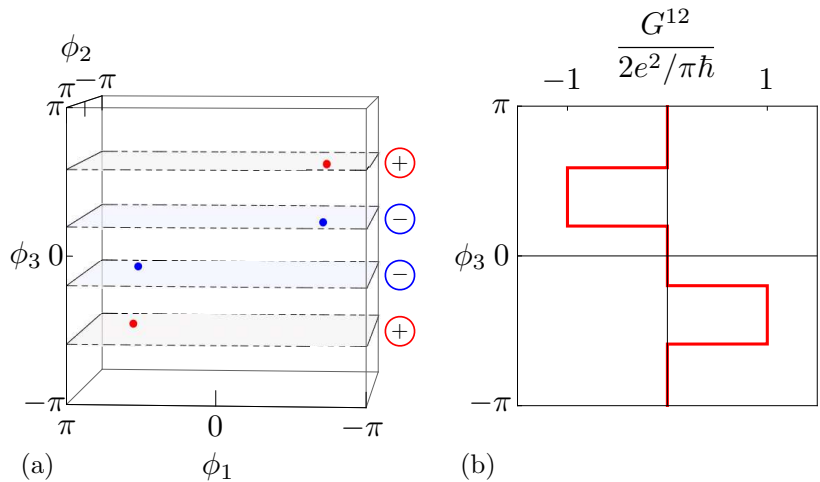}
\caption{\textbf{Topological characterization of the 4-terminal junction for the single-channel case.} a) Position of the four Weyl points in the space of $\phi_{1,2,3}$ of the single-channel 4-terminal junction, the colour code indicating the respective charge. b) The resulting transconductance $G^{12}$ indicating the Chern number, as a function of $\phi_3$ for the same single channel junction as in panel a).}\label{fig_chern_single_channel}
\end{figure}

A specific example is shown in Fig.~\ref{fig_chern_single_channel}. The position and charge of the 4 Weyl points  is shown in Fig.~\ref{fig_chern_single_channel}a. Without loss of generality, we fix the phase $\phi_3$ and compute the transconductance $G^{12}$ between voltage-biased contacts 1 and 2. In Fig.~\ref{fig_chern_single_channel}b, one can clearly see that the transconductance increases (decreases) by $2e^2/(\pi\hbar)$ when $\phi_3$ passes through a Weyl point with positive (negative) charge. Interpreting $\phi_3$ as a control parameter rather than a quasimomentum, we thus see that the 2D bandstructure of the system as a function of $\phi_1$ and $\phi_2$ undergoes a topological transition when $\phi_3$ passes through a Weyl point. The transconductance directly measures the Chern number characterizing the corresponding 2D topological phase. Note that the transconductance satisfies the relation $G^{12}(-\phi_3)=-G^{12}(\phi_3)$ due to time-reversal symmetry.

\begin{figure}
\centering
\includegraphics[scale=0.8]{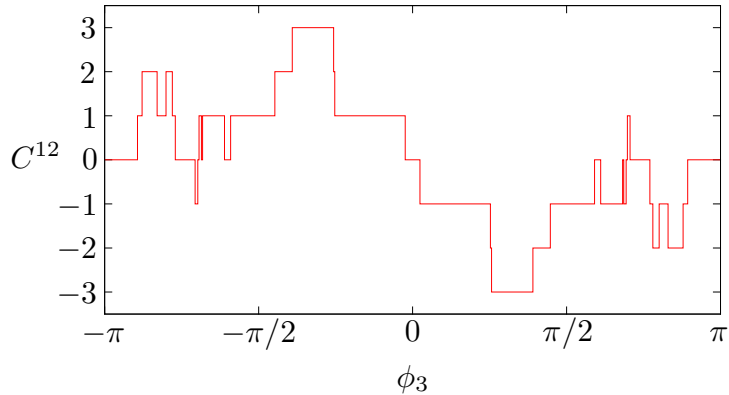}
\caption{\textbf{Topological characterization of the 4-terminal junction for the multi-channel case.} Chern number as a function of $\phi_3$ for a multi-channel 4-terminal junction, where the contacts $1,2,3$, and $0$ contain $12,11,10$, and $9$ channels, respectively. In this particular example, the junction hosts $36$ Weyl points.}\label{fig_chern_multi_channel}
\end{figure}

By randomly generating scattering matrices from the circular orthogonal ensemble~\cite{Mehta2004}, we find that about $5\%$ of scattering matrices give rise to 4 Weyl points (Supplementary Discussion). 
More Weyl points can be obtained in multi-channel junctions where the maximal number of Weyl points is roughly proportional to the number of channels, and the probability to have no Weyl points is small. As a consequence, a greater variety of 2D topological phases with higher Chern numbers can be realized in that case. This is shown in Fig.~\ref{fig_chern_multi_channel} for a multi-channel junction hosting 36 Weyl points, where the maximal Chern number is 3. Recently realized few-channel cross junctions~\cite{Plissard2013} are promising to observe the transconductance in a 4-terminal junction. 

We now turn to 5-terminal junctions. In that case, the Weyl singularities appear as closed loops in the 4D space of phases. The simplest way to visualize them is to consider the additional phase $\phi_4$  as a tuning parameter of the 3D system described by the phases $(\phi_1,\phi_2,\phi_3)$. Tuning $\phi_4$ the Weyl points move, but remain at zero energy. Note that in the 3D subspace, time-reversal symmetry is effectively broken, as for a fixed nonzero $\phi_4$, a Weyl point at $(\phi_1^{(0)},\phi_2^{(0)},\phi_3^{(0)})$ does not have a counterpart at $-(\phi_1^{(0)},\phi_2^{(0)},\phi_3^{(0)})$ anymore. The only constraint at a finite $\phi_4$ is that the number of Weyl points is even. Once two Weyl points with opposite charge meet, they annihilate. Thus their trajectories describe closed loops in the 4D space of all phases.

\begin{figure}
\centering
\includegraphics[scale=0.8]{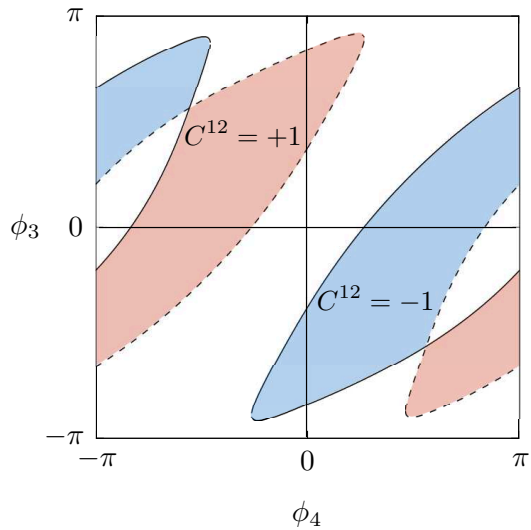}
\caption{\textbf{Topological characterization of a 5-terminal junction, each contact having a single channel.} The colored areas display a nonzero Chern number $C^{12}$. The boundaries of these areas correspond to the projection of the Weyl singularity lines to the $(\phi_{3},\phi_{4})$-plane. In a 3D subspace the curves can be assigned a charge. }\label{fig_5terminal}
\end{figure}

As before, the presence of Weyl singularities may be probed by the transconductance between two voltage-biased terminals, say terminal 1 and 2. As a function of the two other phases $\phi_3$ and $\phi_4$, we now obtain areas with different quantized values of the transconductance, corresponding to different Chern numbers. The boundaries of these areas are given by the projections of the Weyl loops on the $(\phi_3,\phi_4)$-plane. An example is shown in Fig.~\ref{fig_5terminal}. Here time-reversal symmetry of the scattering matrix manifests itself in the relation $G^{12}(-\phi_3,-\phi_4)=-G^{12}(\phi_3,\phi_4)$.

The above considerations are straightforwardly extended to a larger number of terminals. Furthermore, as we saw for the 5-terminal junction, breaking time-reversal invariance does not lift the topological protection of the Weyl points at zero energy, as long as spin degeneracy is preserved. When spin degeneracy is lifted due to a Zeeman field or due to spin-orbit interactions, we expect the Weyl points to shift away from zero energy while remaining stable. The possibility of realizing zero-energy states in 3-terminal junctions with strong spin-orbit interaction was studied in Ref.~\cite{vanHeck2014}. 

To summarize, we predict the existence of topological Weyl singularities in the ABS spectrum of $n$-terminal superconducting junctions with $n\geq4$. These Weyl singularities manifest themselves in a quantized transconductance between two voltage-biased contacts, when the remaining phases are tuned away from the singularities. Tuning the system through a Weyl singularity, the conductance displays a step, signaling a topological transition of the 2D system described by the two phases of the voltage-biased contacts. As with all topological properties, this signature is robust provided the system remains in its ground state. Thus, the quantized transconductance should be easily accessible experimentally at low temperatures and voltages. 

\section*{Acknowledgements}
This work has been supported by the Nanosciences Foundation in Grenoble, in the frame of its Chair of Excellence program. MH and JSM acknowledge support through grants No. ANR-11-JS04-003-01 and No. ANR-12-BS04-0016-03, and an EU-FP7 Marie Curie IRG. 




\end{document}


\title{Supplementary information\\\vspace{1cm}Multi-terminal Josephson junctions as topological materials} 

\author{Roman-Pascal~Riwar, Manuel~Houzet, Julia~S.~Meyer}
\affiliation{Univ. Grenoble Alpes, INAC-SPSMS, F-38000 Grenoble, France,\\
CEA, INAC-SPSMS, F-38000 Grenoble, France
}
\author{Yuli~V.~Nazarov}
\affiliation{Kavli Institute of NanoScience, Delft University of Technology,\\ Lorentzweg 1, NL-2628 CJ, Delft, The Netherlands.}




\maketitle 


\section{Derivation of the Weyl Hamiltonian}

In the following, we provide the derivation of the Weyl Hamiltonian, Eq.~(2) of the main text, and associated properties of the Andreev bound state (ABS) spectrum.

The topological properties of multi-terminal junctions result from the appearance of Weyl singularities in the ABS spectrum. The ABS for an $n$-terminal superconducting junction are determined through~\cite{Beenakker1991}
\begin{equation}
|\psi\rangle=S_N S_A |\psi\rangle\ ,\quad\text{with}\quad S_N=\left(\begin{array}{cc}
S & 0\\
0 & S^{*}
\end{array}\right) \quad\text{and}\quad S_A=\left(\begin{array}{cc}
0 & e^{i\phi}\\
e^{-i\phi} & 0
\end{array}\right)\, e^{-i\chi}\ .
\end{equation}
Here, the two-by-two structure indicates the Nambu space of electrons and holes. The scattering matrix $S_N$ describes the normal metal scattering region, where $S$ and $S^*$ provide the scattering amplitudes for electrons and holes, respectively. The matrix $S_A$ accounts for the phase acquired in the Andreev reflection processes, where $\chi = \arccos(E/\Delta)$ and $e^{i \hat{\phi}}$ is a diagonal matrix that assigns to each channel the phase of the corresponding terminal. Note that the wave function $|\psi\rangle$ is written in terms of the outgoing states
in a given spin sector $\sigma=\uparrow,\downarrow$. 

To determine the ABS energy bands, it is sufficient to reduce the above equation to the electron subspace, where we find the determinant condition for the bound state energies $E$ as
\begin{equation}\label{eq_Beenakker}
{\rm det} \left[ 1 - e^{-2i\chi} A(\hat{\phi})\right]=0\ ,
\end{equation}
with the unitary matrix $A(\hat{\phi})=S e^{i \hat{\phi}} S^* e^{-i \hat{\phi}}$. The matrix $A$ possesses the particle-hole symmetry $A^*=U^\dagger A U$, where $U(\hat{\phi})=e^{i\hat{\phi}} S^T$. This implies that the eigenvalues of $A$ come in pairs~\cite{Note1}, $e^{\pm i a_k}$, corresponding to energies $\pm E_k$, where $E_k=\Delta\cos(a_k/2)$ with $0\leq a_k\leq\pi$.
We assign eigenvectors $|\Psi_k^+\rangle$ and $|\Psi_k^-\rangle=U|\Psi^+_k\rangle^*$ to the pair of eigenvalues $e^{\pm i\alpha_k}$, respectively.

If there is a zero-energy solution at $\hat{\phi}^{(0)}$, the spectrum of the unitary matrix $A(\hat{\phi}^{(0)})$ has a doubly degenerate eigenvalue $-1$. The corresponding orthogonal eigenvectors are given as $|a^+\rangle=|\Psi^+_0(\hat{\phi}^{(0)})\rangle$ and $|a^-\rangle=|\Psi^-_0(\hat{\phi}^{(0)})\rangle$. 
In the vicinity of the singularity at $\hat{\phi}^{(0)}$, we expand the determinant equation~\eqref{eq_Beenakker} for small $\delta\hat{\phi}=\hat{\phi}-\hat{\phi}^{(0)}\ll 1$ and $E\ll\Delta$. Up to first order, we find $e^{i2\chi}\approx-1+2iE/\Delta$ and
\begin{equation}
A(\hat{\phi})\approx A(\hat{\phi}^{(0)})+ i S \delta\hat{\phi} S^\dagger A(\hat{\phi}^{(0)})- i A(\hat{\phi}^{(0)}) \delta\hat{\phi}\ .
\end{equation}
Projecting Eq.~\eqref{eq_Beenakker} onto the subspace defined by $|a^+\rangle$ and $|a^-\rangle$, and keeping only the lowest-order terms, we find the determinant equation
\begin{equation}
\det\left[E-H_\text{W}\right]=0\ .
\end{equation}
It defines the eigenvalue problem for the lowest band, described by the two-by-two Weyl Hamiltonian,
\begin{equation}\label{eq_H_Weyl}
H_\text{W}=\frac{\Delta}{2}\left(\begin{array}{cc}
\left\langle a^{+}\right|\delta\hat{\phi}-S\delta\hat{\phi}S^{\dagger}\left|a^{+}\right\rangle  & \left\langle a^{+}\right|\delta\hat{\phi}-S\delta\hat{\phi}S^{\dagger}\left|a^{-}\right\rangle \\
\left\langle a^{-}\right|\delta\hat{\phi}-S\delta\hat{\phi}S^{\dagger}\left|a^{+}\right\rangle  & \left\langle a^{-}\right|\delta\hat{\phi}-S\delta\hat{\phi}S^{\dagger}\left|a^{-}\right\rangle 
\end{array}\right)\ .
\end{equation}
Using the particle-hole symmetry $|a^-\rangle=U(\hat{\phi}^{(0)})|a^+\rangle^*$, Eq.~\eqref{eq_H_Weyl} may be simplified to
\begin{equation}
H_\text{W}=\sum_{i=x,y,z}h_i\,\tau_i\ ,
\end{equation}
where $\tau_{x,y,z}$ are Pauli matrices in the basis of $\{|a^+\rangle,|a^-\rangle\}$, and the real prefactors $h_i$ take the form
$h_x+ih_y=\Delta \langle\delta\hat{\phi}\rangle_{-+}$ and $h_z=\frac{\Delta}{2}\left(\langle\delta\hat{\phi}\rangle_{++}-\langle\delta\hat{\phi}\rangle_{--}\right)$. As a consequence, the energy spectrum is linear in $\delta\hat{\phi}$ and reads $E=\pm\sqrt{h_x^2+h_y^2+h_z^2}$. We see that 3 independent parameters are necessary to tune the energy to zero. Thus, the Weyl singularities in general appear as points in the space of three independent phases of the 4-terminal junction, and as curves in the space of four independent phases of the 5-terminal junction.

In a given 3D subspace, one may assign a charge to each Weyl point. The sign of a Weyl charge is most conveniently defined by rewriting the Weyl Hamiltonian in matrix form, $H_\text{W}=\sum_{\alpha i}\delta\phi_\alpha M_{\alpha i}\tau_i$ with $M_{\alpha i}=\partial_{\delta\phi_\alpha}h_i$. Then, in a subspace of three independent phases, $M$ is a square matrix, and the sign of the charge is given by the sign of the determinant, $\det[M_{\alpha i}]$.

While particle-hole symmetry is local in $\hat{\phi}$, time-reversal symmetry links $\hat{\phi}$ and $-\hat{\phi}$. As a consequence, a junction with time-reversal symmetry gives rise to pairs of Weyl singularities at $\hat{\phi}^{(0)}$ and $-\hat{\phi}^{(0)}$. Indeed, using $S^T=S$, we find the relation $A(-\hat{\phi})=e^{-i\hat{\phi}}A^\dagger(\hat{\phi}) e^{i\hat{\phi}}$. 
Using this relation, we find that the Weyl Hamiltonians near $\hat{\phi}^{(0)}$ and $-\hat{\phi}^{(0)}$ take the same form and hence, in a given 3D subspace, the corresponding Weyl points have the same charge.

We note that there is the possibility of Weyl singularities at finite energy. Due to particle-hole symmetry, they appear in pairs at energies $\pm E$ with the same charge. Because of particle-hole symmetry and spin degeneracy, solutions at energy $-E$ can be ascribed to solutions at energy $E$ in the opposite spin sector~\cite{Note2}. Thus, zero-energy Weyl points are singly degenerate, while finite energy Weyl points are doubly degenerate. As the finite energy Weyl singularities do not affect the ground state Chern numbers used to characterize the system, we do not discuss them any further. Note that the outgoing wavefunctions in particle-hole space used to compute the Chern numbers are given as $|\psi_k^\pm\rangle=(|\Psi_k^\pm\rangle,e^{-i\chi}U^\dagger|\Psi_k^\pm\rangle)^T$. 
Furthermore, $|\psi_{k\sigma}\rangle$ is identified with $|\psi_k^+\rangle$ in the corresponding spin sector $\sigma$.

\section{Derivation of the current}

In the following, we establish the connection between the current and the Berry curvature for phases that change slowly in time.

The current operator through lead $\alpha$ is defined as
\begin{equation}
\hat{I}_\alpha=2e\frac{\partial \hat{H}}{\partial \phi_\alpha}\ ,
\end{equation}
where $\hat H$ is the Bogoliubov-de Gennes Hamiltonian.
In order to calculate its expectation value for time-dependent phases $\hat{\phi}(t)$, we introduce the basis of instantaneous wave functions of the time-dependent Bogoliubov-de Gennes Hamiltonian $\hat{H}(t)$, such that $E_k(t)|\varphi_{k\sigma}(t)\rangle=\hat{H}(t)|\varphi_{k\sigma}(t)\rangle$. Solving the time-dependent equation $i\hbar|\dot{\varphi}\rangle=\hat{H}(t)|\varphi\rangle$ in that basis, up to first order in phase velocity $\dot{\phi}$, we obtain the current contribution from state $k$ with spin $\sigma$ as 
\begin{equation}\label{eq:current-state-k}
I_{\alpha k}(t)\approx2e\left[\frac{1}{\hbar}\frac{\partial E_k(t)}{\partial\phi_\alpha}-i\frac{\partial\left\langle \varphi_{k\sigma}(t)\right|}{\partial\phi_{\alpha}}|\dot{\varphi}_{k\sigma}(t)\rangle+i\langle \dot{\varphi}_{k\sigma}(t)|\frac{\partial\left|\varphi_{k\sigma}(t)\right\rangle }{\partial\phi_{\alpha}}\right]\ .
\end{equation}
Note that $I_{\alpha k}$ does not depend on spin.
The first term corresponds to the adiabatic supercurrent. Introducing the Berry curvature $B_k^{\alpha\beta}=-2\text{Im}\left[\partial_{\phi_\alpha}\langle\varphi_{k\sigma}|\partial_{\phi_\beta}|\varphi_{k\sigma}\rangle\right]$, the second term reads $-2e\sum_{\beta}\dot{\phi}_\beta B_k^{\alpha\beta}$. 
The many-body expectation value of the current is computed as
\begin{equation}\label{eq_current}
I_\alpha(t)=\sum_{k\sigma}I_{\alpha k}(t)\left[n_{k\sigma}-\frac{1}{2}\right]\ ,
\end{equation}
where $n_{k\sigma}=0,1$ is the occupation of the ABS $k$ with spin $\sigma$.

Equations~\eqref{eq:current-state-k} and \eqref{eq_current} can be used to compute the adiabatic correction to the current, if the energy spectrum is discrete and the occupations of each state do not change with time. For a few-channel junction, the level spacing between the ABS scales as $\sim\!\Delta$, thus providing the adiabaticity condition $\hbar\dot{\phi}\ll\Delta$. On the other hand, defining the current contribution of the states in the continuum above the gap, as well as of ABS that reach the gap edge at some values of the phases, is problematic. In particular, adiabaticity would be broken at any phase velocity for the states in the continuum, while ABS that reach the gap edge may lead to stricter adiabaticity conditions (see next section). In the following, we argue that those states do not contribute to the topological properties considered in our work.

Indeed, multi-terminal junctions are known to be topologically trivial in certain cases. For instance, let us consider the case where terminals $\alpha$ and $\beta$ are voltage biased. When all the other phases $\phi_\gamma$ ($\gamma\neq\alpha,\beta$) are set to zero, the junction is effectively a 3-terminal junction which is topologically trivial. In that case, the associated Chern number $C^{\alpha\beta}$ vanishes. A change of $C^{\alpha\beta}$, corresponding to a topological transition, may occur when a state crosses the Fermi level as the phases $\phi_\gamma$ are varied. As the energy spectrum is gapped, this can only occur for the lowest energy states. Thus, only those states contribute to the topological properties, and in the following we only retain their contribution in Eqs.~\eqref{eq:current-state-k} and \eqref{eq_current}. This establishes the regime of validity for the relation between the transconductance and Chern number given in the main text, see Eq.~(5).  To compute the Chern numbers, we use the outgoing wavefunctions $|\psi_{k\sigma}\rangle$ defined in the previous section~\cite{Note3}.

\section{Comment on Andreev bound states reaching the gap edge}

Let us return to the possibility of the ABS with the highest energy reaching the gap edge. Indeed we find that, for a system with an even number of total channels, this occurs along $(n-1)$-dimensional subspaces in the space of $n$ independent phases~\cite{Note4}.

An argument for this fact can be given in analogy with the derivation of the Weyl Hamiltonian. Namely, an ABS reaching the gap edge at some $\hat{\phi}_\text{g}$ requires that the unitary matrix $A(\hat{\phi}_\text{g})$ has a doubly degenerate eigenvalue $+1$.
We denote the corresponding orthogonal eigenvectors $|b^+\rangle$ and $|b^-\rangle=U(\hat \phi_g)|b^+\rangle^*$. Expanding Eq.~\eqref{eq_Beenakker} for small $\delta\hat{\phi}=\hat{\phi}-\hat{\phi}_\text{g}$, we can reduce the determinant equation to
\begin{equation}
\det\left[{\rm sign}(E)\sqrt{\frac{8(\Delta-|E|)}{\Delta}}+\left(\langle\delta\hat{\phi}\rangle_{++}-\langle\delta\hat{\phi}\rangle_{--}\right)\tilde{\tau}_z\right]=0\ ,
\end{equation}
where $\tilde{\tau}_z$ is a Pauli matrix in the basis $\{|b^+\rangle,|b^-\rangle\}$. 

Thus, the solution for the ABS energies reads
\begin{equation}
E\approx\pm \Delta\left[1- \frac 1 8 \left(\langle\delta\hat{\phi}\rangle_{++}-\langle\delta\hat{\phi}\rangle_{--}\right)^2\right]\ .
\label{eq-theta}
\end{equation}
Furthermore, we see that ABS wave function has a discontinuity at $\hat{\phi}_\text{g}$. Namely, the wave function of the state at positive energy is given by $|b^+\rangle$ for $\langle\delta\hat{\phi}\rangle_{++}-\langle\delta\hat{\phi}\rangle_{--}<0$ and by $|b^-\rangle$ for $\langle\delta\hat{\phi}\rangle_{++}-\langle\delta\hat{\phi}\rangle_{--}>0$, and vice versa for the states at negative energies.

\begin{figure}
\centering
\includegraphics[scale=0.8]{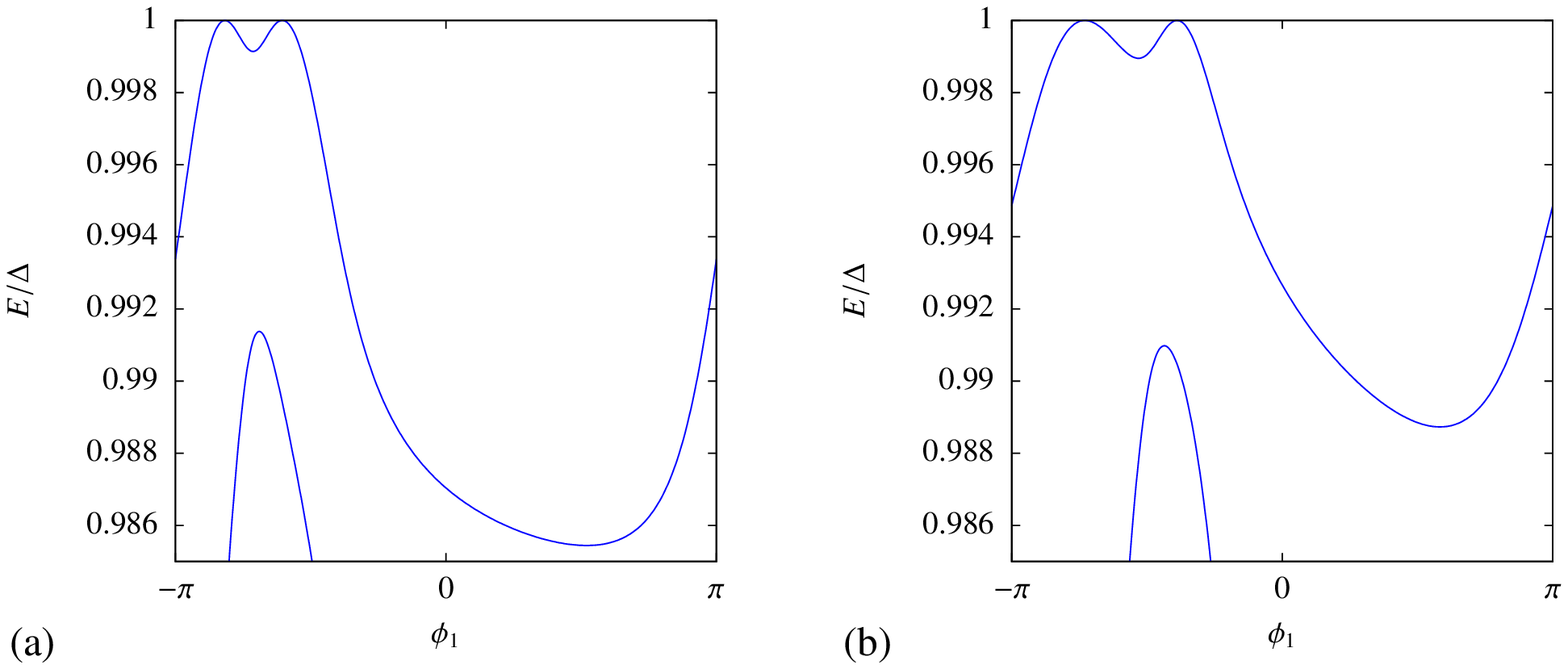}
\caption{\textbf{Zoom on the highest ABS of the spectra shown in Figs.~1b and 1c in the main text.} The gap edge touchings are clearly visible.} 
\label{fig_gapedge}
\end{figure}

In contrast to the discussion for the zero-energy Weyl points, here a single parameter is sufficient to tune the ABS energy to the gap edge. Therefore, the highest ABS reaches the gap edge along $(n-1)$-dimensional subspaces in the space of $n$ independent phases. This is visible in Figs.~1b and 1c in the main text. A zoom is shown in Fig.~\ref{fig_gapedge}.
  
When sweeping two phases, the curves where the highest ABS reaches the continuum cannot be avoided. If this state was occupied, this would break adiabaticity even for arbitrary slow driving. (Introducing a finite energy-dependence of the scattering matrix or spin-orbit interactions may create a finite level repulsion between the highest ABS and the continuum and, thus, restore adiabaticity, possibly under stricter conditions. This would need further investigation.) However, as far as the topological characterization of the junction is concerned, we came to the conclusion that this peculiarity should not change our results.

\section{Estimation of the current noise}

The topological signal of interest is a small dc contribution on top of a large ac adiabatic current. This makes it important to provide an estimate for the measurement time $T_0$ required to sufficiently average the current signal. In particular, the lower bound for  $T_0$  is provided by $S_I/|\overline{I}|^2$, where $\overline{I}$ denotes the dc current signal and $S_I$ is the low-frequency current noise. The dominant contribution to $S_I$ stems from fluctuations of the phases due to voltage noise and will be estimated in the following.

We define the current noise as
\begin{equation}
\label{eq:SI}
S_I=\int d\tau \left\langle I(t) I(t+\tau)\right\rangle\ ,
\end{equation}
where the brackets $\left\langle\ldots\right\rangle$ denote the expectation value with respect to the statistical ensemble. As the adiabatic supercurrent, i.e., the first term in Eq.~\eqref{eq_current}, is much larger than the correction proportional to the Berry phase, it dominates the noise.
We exploit the fact that this current is periodic in both phases $\phi_\alpha$ and $\phi_\beta$, such that it may be expanded in Fourier harmonics,
\begin{equation}
\label{eq:harmonics}
I(\phi_{\alpha},\phi_{\beta}) = \sum_{n_\alpha,n_\beta=-\infty}^\infty I_{n_\alpha,n_\beta} e^{i (n_\alpha \phi_\alpha  + n_\beta \phi_\beta )}\ .
\end{equation}
Substituting Eq.~\eqref{eq:harmonics} into Eq.~\eqref{eq:SI}, we obtain
\begin{equation}
\label{eq_noise}
S_I = \int d\tau \sum_{n_\alpha,n_\beta} |I_{n_\alpha,n_\beta}|^2 \langle e^{i n_\alpha[\phi_\alpha(t)-\phi_\alpha(t+\tau)]}\rangle \langle e^{i n_\beta[\phi_\beta(t)-\phi_\beta(t+\tau)]}\rangle\ ,
\end{equation}
as the two phases fluctuate independently and the ac contribution averages to zero.

In order to evaluate the averages, we assume white noise in the voltage sources, described by
\begin{equation}
\label{eq:correlators}
\frac{2e^2}{\hbar^2}\left\langle\delta V_\alpha(t)\delta V_\alpha(t')\right\rangle=\Gamma_\alpha\delta(t-t')\ ,
\end{equation}
with the rate $\Gamma_\alpha$ associated to the voltage noise in lead $\alpha$. 
Using $\phi_\alpha(t+\tau)-\phi_\alpha(t)=\omega_\alpha \tau+(2e/\hbar)\int_t^{t+\tau}ds\, \delta V_\alpha(s)$, 
where  
$\omega_\alpha \equiv 2eV_\alpha/\hbar$, we find
\begin{equation}
\langle e^{i n_\alpha[\phi_\alpha(t)-\phi_\alpha(t+\tau)]}\rangle = e^{- i n_\alpha \omega_\alpha \tau - \Gamma_\alpha n_\alpha^2 |\tau|} \ ,
\end{equation}
and subsequently
\begin{equation}
\label{eq:noise2}
S_I = 2 \sum_{n_\alpha,n_\beta} |I_{n_\alpha,n_\beta}|^2 \frac{\Gamma_\alpha n_\alpha^2 +\Gamma_\beta n_\beta^2}{(n_\alpha \omega_\alpha + n_\beta \omega_\beta)^2 + (\Gamma_\alpha n_\alpha^2 +\Gamma_\beta n_\beta^2)^2}\ .
\end{equation}
We see that $S_I$ depends on frequencies $\omega\sim\omega_{\alpha,\beta}$ in a quite complex fashion. In the limit of low voltage noise, $\omega\gg\Gamma$ with $\Gamma\sim\Gamma_{\alpha,\beta}$, we have to distinguish between commensurate and incommensurate voltages $V_{\alpha,\beta}$. In the commensurate case, the first term in the denominator of Eq.~\eqref{eq:noise2} vanishes for a pair $(n_\alpha,n_\beta)$, yielding $S_I\propto 1/\Gamma$. By contrast, in the incommensurate case, the first term in the denominator of Eq.~\eqref{eq:noise2} dominates, yielding a much weaker noise, $S_I\propto \Gamma/\omega^2$. Thus, the current noise strongly varies with $\omega$. Averaging over a window of width $\sim\omega$ yields the average noise $\overline{S_I}\propto 1/\omega$. On the other hand, in the noisy regime $\omega\ll\Gamma$, we find that $S_I\propto 1/\Gamma$, irrespective of the voltages being commensurate or not. Thus, the current noise is actually weaker than the average current noise in the opposite regime. Furthermore, incommensurability is no longer required for averaging, as the strong phase fluctuations take care of covering the entire unit cell.

This brings us to the conclusion that the strong voltage noise regime $\omega\ll\Gamma$ may be favourable for averaging -- provided, of course, that the noise is still in the limit $\hbar\Gamma\ll\Delta$. To estimate the measurement time, we use $I_{n_\alpha,n_\beta}\sim e\Delta/\hbar$ and $\bar I\sim e^2V/\hbar$ with $V\sim V_{\alpha,\beta}$, together with the estimate for $S_I$ found above,  to obtain
\begin{equation}
T_0>\frac{\Delta^2}{(eV)^2}\frac{1}{\Gamma}\ .
\end{equation}

\section{Occurence of Weyl points for 4-terminal junctions}

Here, we estimate how often random scattering matrices give rise to Weyl points at zero energy. 

As discussed in the first section of the supplementary, such Weyl singularities emerge when the matrix $A$ has the eigenvalue $-1$. Whether this occurs indeed
depends on the properties of the junction, encoded in the scattering matrix $S$.
We performed a numerical analysis for 4-terminal junctions, both in the single-channel and multi-channel cases. For this purpose, we generated symmetric scattering matrices according to the circular orthogonal ensemble using the following recipe~\cite{Mehta2004}: We randomly generated hermitian matrices $\mathcal{H}$ from a Gaussian ensemble, and numerically diagonalized them, $\mathcal{H}=\mathcal{U}^\dagger\mathcal{D} \mathcal{U}$, where $\mathcal{U}$ is a unitary matrix and $\mathcal{D}$ is a real diagonal matrix. A random symmetric scattering matrix was then generated as $S=\mathcal{U}^T\mathcal{U}$. For each $S$, we numerically checked the existence of zeros of the function
$\left|\det[1+A(\hat{\phi})]\right|$ in the space of phases. 

In the single-channel case, we ran the check for 965 randomly generated matrices $S$, out of which 46 gave rise to zero-energy solutions. Thus, we found that a total of roughly $5\%$ of all scattering matrices yield Weyl points, while the remaining $95\%$ provide a trivial junction.

\begin{figure}
\centering
\includegraphics[scale=0.8]{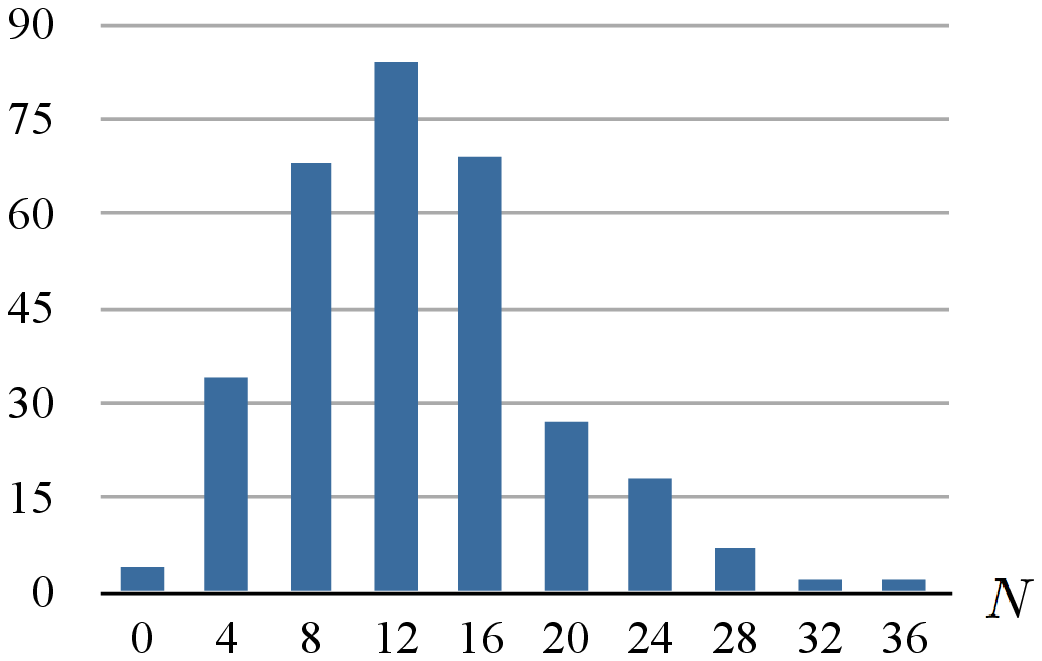}
\caption{\textbf{Histogram displaying the occurrence of random scattering matrices yielding $N$ Weyl points for the four-terminal multi-channel junction.}} 
\label{fig_Weyl_histogram}
\end{figure}

When increasing the number of channels in each terminal, we found that the maximal number of Weyl points scales with the number of channels, and that the probability of a junction without Weyl points decreases significantly.  A total of 324 random scattering matrices were generated for a junction with four terminals, where the terminals have 12,11,10, and 9 channels, respectively. In Fig.~\ref{fig_Weyl_histogram} we show the histogram displaying the occurrence of randomly generated scattering matrices that provide $N$ zero-energy Weyl points. Only 4 scattering matrices gave rise to a junction without zero-energy Weyl points. Note that our algorithm has a small, but finite probability to miss some zeros. As a consequence, while the number of Weyl points has to be a multiple of $4$ because of time-reversal symmetry (see main text), the algorithm misidentifies a few cases as having an odd multiple of 2 Weyl points.
